# *Ab initio* study of elastic and electronic properties of cubic thorium pnictides Th*Pn* and Th$_3$*Pn*$_4$ (*Pn* = P, As, and Sb)

I.R. Shein, * A.L. Ivanovskii

*Institute of Solid State Chemistry, Ural Branch of the Russian Academy of Sciences, 620990, Ekaterinburg, Russia*

A B S T R A C T

Full-potential linearized augmented plane-wave method with the generalized gradient approximation for the exchange-correlation potential was applied for comparative study of elastic and electronic properties of six cubic thorium pnictides Th*Pn* and Th$_3$*Pn*$_4$, where *Pn* = P, As, and Sb. Optimized lattice parameters, theoretical density, independent elastic constants ($C_{ij}$), bulk moduli (*B*), shear moduli (*G*), Young's moduli (*Y*), and Poisson's ratio (ν) were obtained for the first time and analyzed in comparison with available theoretical and experimental data. The electronic band structures, total and partial densities of states for all Th*Pn* and Th$_3$*Pn*$_4$ phases were examined systematically. Moreover, the inter-atomic bonding pictures in thorium pnictides, as well as the relative stability of Th*Pn versus* Th$_3$*Pn*$_4$ phases were discussed.

*Keywords:* Cubic thorium pnictides Th*Pn*, Th$_3$*Pn*$_4$; Electronic, Elastic Properties; Chemical bonding; *ab initio* calculations

----------------

* Corresponding author: *E-mail address:* shein@ihim.uran.ru (I.R. Shein)

# 1. Introduction

Binary compounds formed by thorium with *sp* elements have attracted much attention owing to their unique physical and chemical properties making them interesting from the fundamental point of view, as well as for a variety of technical applications, for example as alternative fertile materials to be used in nuclear breeder systems *etc.*, see [1-5].

Thorium pnictides belong to this important group of materials. For example, the refractory ThP demonstrates good heat conductivity and chemical stability and holds promise as nuclear fuel for special reactors; $Th_3As_4$ may have desirable properties for thermoelectric applications *etc.*, see [5].

Thorium reacts with pnictogens (*Pn* = P, As, and Sb) to form two main phases: *mono-* and *tetra-*pnictides. Thorium *mono-*pnictides (Th*Pn*) crystallize in the cubic *B*1-like structure, where *Pn* atoms occupy the octahedral interstitial sites in the face-centered cubic thorium sublattice. Thorium *tetra-*pnictides $Th_3Pn_4$ crystallize in a cubic structure of the $Th_3P_4$ type, which is prototypical for many compounds of transition metals, lanthanides and actinides, which also display interesting physical properties [1,6-8]. Note that some other so-called *poly-*pnictides (pnictogen-rich phases Th$Pn_{z>4/3}$ such as Th$Pn_2$, $Th_2Pn_{11}$ or Th$Pn_7$) are also known [5].

Despite a number of interesting results on the physical properties of Th*Pn* and $Th_3Pn_4$ materials, the data on their mechanical behavior and electronic properties, to our knowledge, are limited to several early works devoted mainly to *mono-*pnictides Th*Pn* [5,9-17].

In this paper, in order to get a systematic insight into the elastic and electronic properties of these materials, a comparative first-principle study of six cubic thorium pnictides Th*Pn* and $Th_3Pn_4$ (*Pn* = P, As and Sb) was performed using the FLAPW method within the generalized gradient approximation (GGA) for the exchange-correlation potential. This choice allowed us to compare the above properties of these related phases as a function of: (i) the pnictogen type (*Pn* = P, As, and Sb) and (ii) the ratio of Th/*Pn* atoms (Th/*Pn* =1 *versus* Th/*Pn* =3/4, *i.e.* Th*Pn* ↔ Th$_3$*Pn*$_4$).

As a result, we evaluated and systematically analyzed a set of physical parameters of the above-mentioned Th*Pn* and Th$_3$*Pn*$_4$ such as optimized lattice parameters, density, elastic constants ($C_{ij}$), bulk (*B*), shear (*G*), and Young's moduli (*Y*), Poisson's ratio (*v*), tetragonal shear moduli (*G'*, in GPa), Cauchy's pressure (*CP*), and Zener's anisotropy index.

In addition, the electronic band structure and intra-atomic bonding picture for these materials were also examined, and their general trends were discussed.

## 2. Computational details

Thorium *mono*-pnictides (Th*Pn*) adopt a face-centered cubic NaCl-type structure (space group *Fm3m*, No. 225); the thorium atoms are placed in 4(*a*) and *Pn* atoms – in 4(*b*) positions. Each Th and *Pn* atom is surrounded by six *Pn* or Th atoms, respectively, forming a regular octahedrons.

Thorium *tetra*-pnictides (Th$_3$*Pn*$_4$) adopt a body-centered cubic Th$_3$P$_4$-type structure (space group *I43d*, No 220, *Z* = 4); the thorium atoms are placed in 12(*a*) and *Pn* atoms – in 16(*c*) positions. Each Th atom is surrounded by 8 *Pn* atoms, whereas each *Pn* atom is surrounded by six Th atoms forming a distorted octahedrons. The mentioned Th*Pn* and Th$_3$*Pn*$_4$ structures are illustrated in Fig. 1.

The calculations of the all mentioned Th pnictides were carried out by means of the full-potential method with mixed basis APW+lo (LAPW) implemented in the WIEN2k suite of programs [18]. The generalized gradient correction (GGA) to exchange-correlation potential of Perdew, Burke and Ernzerhof [19] was used. The sphere radii were chosen 2.5 a.u. for Th and 2.0 a.u. for pnictogens. The starting configurations were: [Rn]($6p^6 6d^2 7s^2 7p^0 5f^0$) for thorium, [Ne]($3s^2 3p^3$) for P, [Ar]($3d^{10} 4s^2 4p^3$) for As, and [Kr]($4d^{10} 5s^2 5p^3$) for Sb. The maximum value for partial waves used inside atomic spheres was $l = 12$ and the maximum value for partial waves used in the computation of muffin-tin matrix elements was $l = 4$. The plane-wave expansion with RMT ×KMAX equal to 7 and k sampling with 12×12×12 k-points mesh in the Brillouin zone were used. Relativistic effects were taken into account within the scalar-relativistic approximation. The self-consistent calculations were considered to converge when the difference in the total energy of the crystal did not exceed 0.001 mRy as calculated at consecutive steps. Other details of the calculations are described in Ref. [20].

## 3. Results and discussion

### 3.1. Lattice constants and density

The equilibrium lattice constants ($a$) for Th$Pn$ and Th$_3Pn_4$ phases were calculated under constraining of cubic symmetry. The results are given in Table 1, and agree reasonably with the available experimental data. As can be seen, the predicted lattice constants differ from the experimental values by no more than ~ 0.5–1.5%, which is a typical overestimation of lattice constants for methods employing GGA. In turn, $a$ of Th$Pn$ and Th$_3Pn_4$ phases increases in the following sequence: $a$(ThP) < $a$(ThAs) < $a$(ThSb) and $a$(Th$_3$P$_4$) < $a$(Th$_3$As$_4$) < $a$(Th$_3$Sb$_4$). This result can be easily explained by considering the anionic radii $R$ of pnictogens: $R$(P) = 1.920 Å < $R$(As) = 1.993 Å < $R$(Sb) = 2.171 Å.

Next, the calculated cell volumes were used for estimation of theoretical density ($\rho^{theor}$) of the considered phases. The data obtained (Table 1) showed that the density of these materials decreases in the following sequence: $\rho^{theor}$(ThAs) > $\rho^{theor}$(ThSb) > $\rho^{theor}$(ThP) and $\rho^{theor}$(Th$_3$As$_4$) > $\rho^{theor}$(Th$_3$Sb$_4$) > $\rho^{theor}$(Th$_3$P$_4$). Thus, the As-containing phases have higher density than P- or Sb-containing phases; at the same time *tetra*-pnictides (excluding Th$_3$As$_4$) are lighter than the corresponding *mono*-pnictides.

*3.2. Elastic parameters*

At first, we have calculated elastic constants $C_{ij}$ for Th*Pn* and Th$_3$*Pn*$_4$, Table 1. These three independent elastic constants in the cubic symmetry ($C_{11}$, $C_{12}$, and $C_{44}$) were estimated by calculating the stress tensors on strains applied to equilibrium structure. The data obtained allow us to make the following conclusions.

All of the elastic constants $C_{ij}$ for Th*Pn* and Th$_3$*Pn*$_4$ phases are positive and satisfy the generalized criteria [21] for mechanically stable cubic crystals: $C_{11} > 0$; $C_{44} > 0$, $(C_{11} - C_{12}) > 0$; $(C_{11} + 2C_{12}) > 0$.

Then, the Voigt-Reuss-Hill (VRH) approximation [22] was used to obtain the bulk moduli (*B*) and shear moduli (*G*). It was found (Table 1) that the bulk moduli of the cubic Th*Pn* and Th$_3$*Pn*$_4$ decrease in the sequence: $B$(ThP) > $B$(ThAs) > $B$(ThSb) and $B$(Th$_3$P$_4$) > $B$(Th$_3$As$_4$) > $B$(Th$_3$Sb$_4$) – in agreement with the well-known relationship [23] between *B* and the lattice constants (cell volume $V_o$, as $B_o \sim k/V_o$), whereas the compressibility of Th pnictides ($\beta = 1/B$) changes in the inverse sequence. On the other hand, for the cubic Th*Pn* and Th$_3$*Pn*$_4$ phases with the same pnictogens their bulk moduli and compressibility are comparable, Table 1.

The Young's modulus is defined as the ratio between stress and strain and is used to provide a measure of stiffness, *i.e.* the larger is the value of *Y*, the stiffer is the material. In our case $\{Y(\text{Th}_3\text{P}_4) > Y(\text{ThP})\} > \{Y(\text{Th}_3\text{As}_4) > Y(\text{ThAs})\} > \{Y(\text{Th}_3\text{Sb}_4) > Y(\text{ThSb})\}$, Table 1.

The calculated shear moduli *G* (as well as the tetragonal shear moduli *G*' = $(C_{11}-C_{12})/2$) indicate that thorium phospides will possess the maximal resistance to shear deformation as compared with Th arsenide and Th antimonides. Let us note that for all Th*Pn* and Th$_3$*Pn*$_4$ phases $B > G' > G$; this implies that a parameter limiting the stability of these materials is the shear modulus *G*.

The *intrinsic* hardness together with compressibility belongs to the most important mechanical characteristics of materials. However, hardness is a macroscopic concept, which is characterized experimentally by indentation, and thus depends strongly on plastic deformation – unlike the compressibility related to elastic deformation [24,25]. Even though these deformations (plastic and elastic) are fundamentally different, the values of the bulk moduli *B* (which measures the resistance to volume change for a constant shape) and of the shear moduli *G* (which measures the resistance to shape change at a constant volume) are often used [26-28] as preliminary hardness predictors. In our case, both *B* and *G* moduli adopt the maximal values for thorium phosphides, therefore these materials are expected to exhibit the maximal hardness.

Another important mechanical characteristic of materials is their brittle/ductile behavior, which is closely related in particular to their reversible compressive deformation and fracture ability. Here, the widely used malleability measures are the Pugh's indicator (*G/B* ratio) [29] and the Cauchy's pressure [30] $CP = (C_{12}-C_{44})$. As is known empirically, a material behaves in a ductile manner if *G/B* < 0.5, and *vice versa*, if *G/B* > 0.5, a material demonstrates brittleness. As regards Cauchy's pressure, this parameter is also used sometimes to explore chemical bonding geometry since it has negative values for directionally bonded (strong covalent) solids such as diamond, and positive ones for ductile metals [31]. An additional indicator of brittle/ductile behavior follows from the Poisson's ratio *v*: these values for brittle covalent materials are small, whereas for ductile metallic materials *v* is typically 0.33 [26]. Our data reveal that for the majority of the examined Th pnictides *G/B* ~ 0.5 and the values of the Poisson's ratio vary in the interval 0.28-0.34; thus these phases will lie on the brittle/ductile border. The Cauchy's pressure is negative only for ThP (adopting also the maximal values of *B* and *G* moduli), which will therefore contain a considerable covalent contribution to the inter-atomic bonds, see also below.

The elastic anisotropy of crystals is of importance for engineering science since it correlates with the possibility to induce microcracks in materials [32-34]. We have estimated the elastic anisotropy for the examined materials using the so-called Zener's anisotropy index $A = 2C_{44}/(C_{11} - C_{12})$ [34]. For isotropic crystals $A = 1$, while values smaller or greater than unity measure the degree of elastic anisotropy. In our case: (i) the minimal anisotropy is adopted by thorium *tetra*-pnictides Th$_3$*Pn*$_4$ - as compared with the same for thorium *mono*-pnictides Th*Pn*, and (ii) in each series of *tetra*- or *mono*-pnictides the Zener's index decreases as going from phosphide to antimonides. Thus, according to our estimations, Th$_3$P$_4$ will behave as the most isotropic material.

*3.3. Electronic properties*

Figures 2-4 show the band structures and total and atomic-resolved *l*-projected densities of states (DOSs) in Th*Pn* and Th$_3$*Pn*$_4$ phases as calculated for equilibrium geometries. First of all, we found that these Th pnictides are divided into three types: metallic-like *mono*-pnictides, semiconducting (Th$_3$P$_4$ and Th$_3$As$_4$), and semimetallic (Th$_3$Sb$_4$) *tetra*-pnictides.

Let us discuss the common features of the electronic structure of metallic-like Th*Pn* phases using ThP as an example, Figs. 2 and 4. In the valence spectrum of the *mono*-phosphide, the lowest band lying around -10 eV below the Fermi level ($E_F$) arises mainly from P 3*s* states and is separated from the near-Fermi bands by a gap. These bands are located in the energy range from -5.3 eV to $E_F$ and are formed predominantly by P 3*p* and Th (*p,d,f*) states. The corresponding DOSs include two main subbands A and B, Fig. 4. The subband A contains strongly hybridized Th (*p,d,f*)–P 3*p* states, which are responsible for the covalent Th-P bonds, see also below. The topmost part of the valence band (subband B) is made up basically of contributions from Th 6*d*,5*f* states; these states also contribute to the bottom of the conducting band (peak C, Fig. 4), where Th 5*f* states dominate. Thus, as in metallic thorium [1,35,36] and other thorium compounds with light *sp* atoms (H, B, C, N,O) [17,37-42], the Th 5*f* states for ThP are itinerant and partially occupied.

For better understanding of the composition of the near-Fermi bands, the total and orbital decomposed partial DOSs at the Fermi level, $N(E_F)$, are listed in Table 2. It is seen that the main contributions to $N(E_F)$ are from the Th $5f$ and the Th $6d$ states. These data allow us to estimate the Sommerfeld constants ($\gamma$) and the Pauli paramagnetic susceptibility ($\chi$), assuming the free electron model, as: $\gamma = (\pi^2/3)N(E_F)k^2_B$ and $\chi = \mu_B^2 N(E_F)$, Table 2. It is seen that the evaluated $\gamma$ for ThP is in reasonable agreement with the available experimental data, measured at ambient temperature [5]. The common trend is the increase of $\gamma$ and $\chi$ as going from ThP to ThSb. This effect may be related with the increase of the Th $5f$ contributions in $N(E_F)$, see Table 2.

In general, the electronic structures for other metallic-like *mono*-pnictides (ThAs and ThSb) are similar to the same for the above isostructural and isoelectronic ThP, see Figs. 2 and 4. The main differences for Th*Pn* phases with various types of pnictogens are in their bandwidths listed in Table 3.

For *tetra*-pnictides $Th_3Pn_4$, the energy sequence of the main occupied bands (*Pn s* and *Pn p* + Th (*p,d,f*) states) and the itinerant nature of Th $5f$ states are similar to those for the above Th*Pn* phases, Figs. 3 and 4. However, unlike Th*Pn*, the occupied near-Fermi states are formed predominantly by hybridized *Pn p* + Th (*p,d,f*) states with appreciable contributions of *Pn p* orbitals.

The band gap (BG) values are among the most important parameters for the semiconducting materials. The results of our FLAPW-GGA calculations of BGs for $Th_3Pn_4$ are listed in Table 4. According to the obtained data, $Th_3P_4$ and $Th_3As_4$ are indirect-band-gap ($\Delta \rightarrow \Gamma$) type semiconductors with BGs of 0.17 eV and 0.28 eV, respectively, whereas $Th_3Sb_4$ belongs to semimetals. Meanwhile it is well known that first-principles band structure methods using the local density approximation (LDA) and the related generalized approximation (GGA) give very reasonable results for the ground state properties (such as structural parameters, total energies *etc.*), but lead to typical underestimating of the band gap (BG) for semiconducting or insulating materials - at least by 30 ÷ 50% [44]. Thus, the above values of band gaps should be considered as a bottom limit. To improve the theoretical estimations of the band gaps, for Th$_3$Pn$_4$ systems we used also the so-called hybrid exchange-correlation potential in the DFT-GGA+HF form [45] as implemented in the WIEN2k code. From the results presented in Table 4, we can see that for Th$_3$P$_4$ and Th$_3$As$_4$ the DFT-GGA+HF potential leads to an appreciable increase in the theoretical gaps, which are in good agreement with experimental reports. For Th$_3$Sb$_4$, the use of the DFT-GGA+HF potential does not change the semi-metallic picture of the electronic spectrum.

Let us discuss the intra-atomic bonding in Th*Pn* and Th$_3$Pn$_4$ phases. The character of *covalent bonding* in these systems may be well understood from site-projected DOS calculations. As is shown in Fig. 3, *Pn p* and Th (*p,d,f* ) states are strongly hybridized. These covalent Th-*Pn* bonds are clearly visible in Figs. 5 and 6 displaying the charge density map in the (100) plane of B1-like thorium *mono*-phosphide ThP (Fig. 5) and the isoelectronic surface of the charge density in the cubic *tetra*-phosphide Th$_3$P$_4$ (Fig. 6), which are typical also of other examined Th*Pn* and Th$_3$Pn$_4$ phases. In addition, the charge density map depicted in Fig. 5 demonstrates that the *metallic-like* Th-Th interaction occurs owing to delocalized near-Fermi Th *d,f* states (subband B, Fig. 4).

To estimate the amount of electrons redistributed between the Th and pnictogens (i.e. *ionic bonding*), we carried out a Bader [46] analysis. The total charge of an atom (the so-called Bader charge, $Q^B$) as well as the corresponding effective charges ($\Delta Q^{eff}$) are presented in Table 5. These results show that the charge transfer (Th→*Pn*) is much smaller than it is predicted in the idealized ionic model (3 *e* per *Pn* atom) – owing to the above hybridization effects. Namely, for *mono*-pnictides this transfer $\Delta Q^{eff}$(Th→*Pn*) is about 1.5 – 1.7 *e* per atom; in the series ThP → ThAs → ThSb the effective atomic charges decrease. In turn, for Th$_3$Pn$_4$ phases the effective charges ($\Delta Q^{eff}$) per Th atom increase and $\Delta Q^{eff}$ per *Pn* atom decrease as compared with the corresponding thorium *mono*-pnictides. Besides, similarly to the Th*Pn* series, the effective atomic charges in the sequence $Th_3P_4 \rightarrow Th_3As_4 \rightarrow Th_3Sb_4$ decrease, Table 5.

Finally, the values of the total energies of Th pnictides as obtained in our band structure calculations allow us to estimate the relative stability of the corresponding phases, namely Th*Pn* *versus* Th$_3$*Pn*$_4$. For this purpose, the formation energies of *mono*-pnictides (for the formal reactions Th$_3$*Pn*$_4$ + Th → Th*Pn*) were calculated as $E_{form} = E_{tot}(ThPn) - \{1/4 E_{tot}(Th) + 1/4 E_{tot}(Th_3Pn_4)\}$.

The results obtained lead to the following conclusions. Firstly, for all Th pnictides the calculated values of $E_{form}$ are negative; this means that Th*Pn* are more stable than Th$_3$*Pn*$_4$. Secondly, for ThP the estimated $E_{form}$ (-12.66 eV/atom Th, *i.e.* about -291.72 kcal/mol) is in reasonable agreement with the available experimental data (-274 kcal/mol [5]). The energy of formation for ThSb (-11.43 eV/atom Th) is comparable to the latter value, whereas the calculated $E_f$ for ThAs is much smaller (-0.5 eV/atom Th). This indicates that the formation of the mixture of phases (ThAs+Th$_3$As$_4$) *via* various synthetic routes [5] is more probable for the Th-As system than for the systems Th-P and Th-Sb.

## 4. Conclusions

In summary, the first-principles calculations were performed for systematical study of the structural, elastic and electronic properties for the family of six cubic thorium pnictides Th*Pn* and Th$_3$*Pn*$_4$, where *Pn* = P, As and Sb.

The evaluated elastic parameters allow us to conclude that all cubic thorium pnictides are mechanically stable; the parameter limiting their mechanical stability is the shear modulus *G*. Our data reveal the intervals of elastic parameters, which are typical of cubic thorium pnictides. So, the maximal and minimal values of bulk moduli are 111-107 GPa (for Th$_3$P$_4$ and ThP) and 82-79 PGa (for ThSb and Th$_3$Sb$_4$); the shear moduli vary from 58-56

GPa (for Th$_3$P$_4$ and ThP) to 41-28 GPa (for Th$_3$Sb$_4$ and ThSb). Thus, as both *B* and *G* moduli adopt the maximal values for thorium phosphides, and these materials are be expected to have the maximal hardness. Our data reveal also that the majority of the examined Th pnictides will lie on the brittle/ductile border. In addition, according to our estimations, the thorium *tetra*-pnictides Th$_3$*Pn*$_4$ adopt the minimal elastic anisotropy – as compared with the same thorium *mono*-pnictides Th*Pn*, whereas in each series of *tetra*- or *mono*-pnictides the Zener's index decreases as going from phosphide to antimonide. As a result, Th$_3$P$_4$ will behave as the most isotropic material.

Our calculations show that, as distinct from metallic-like *mono*-pnictides, thorium *tetra*-pnictides Th$_3$P$_4$ and Th$_3$As$_4$ are indirect-band-gap semiconductors (with band gaps at about 0.40 eV and 0.44 eV, respectively, as obtained using hybrid DFT-GGA+HF potential), whereas Th$_3$Sb$_4$ behaves as a semimetal. Let us note also that for all thorium pnictides the Th 5*f* states are itinerant and partially occupied. The intra-atomic bonding picture in the cubic thorium pnictides is a combination of covalent, ionic and metallic contributions. Comparable contributions of Th 6*d* and 5*f* states are responsible for the formation of Th–*Pn* and Th–Th bonds, whereas the charge transfer Th→*Pn* determines the ionic component of bonding. Finally, our numerical estimations of the formation energies for Th*Pn versus* Th$_3$*Pn*$_4$ indicate that Th*Pn* phases are more stable than the corresponding Th$_3$*Pn*$_4$ phases.

**Table 1**.

Calculated structural and elastic parameters for cubic thorium pnictides Th*Pn* and Th$_3$*Pn*$_4$ (*Pn* = P, As, and Sb) in comparison with available data. Lattice parameters (*a*, in Å), theoretical density ($\rho^{theor}$, in g/cm$^3$), independent constants ($C_{ij}$, in GPa), bulk moduli (*B*, in GPa), shear moduli (*G*, in GPa), tetragonal shear moduli (*G'*, in GPa), Young's moduli (*Y*, in GPa), Poisson's ratio (ν), Pugh's indicator (*G/B*), Cauchy's pressure (*CP*, in GPa), Zener's anisotropy index (*A*). Available experimental [5,11] and calculated [15] data are given in parentheses.

| Phase | ThP | ThAs | ThSb |
|---|---|---|---|
| *a* | 5.8578 (5.827-5.840 *[a]) | 6.0144 (5.960-5.978 [a]) | 6.3723 (6.318 [a]) |
| $\rho^{theor}$ | 8.691 (8.83-8.77 *[a]) | 9.372 | 9.082 |
| $C_{11}$ | 263.60 (278.77 [c]) | 244.90 (246.24 [c]) | 204.50 (208.51 [c]) |
| $C_{12}$ | 29.60 (46.45 [c]) | 27.60 (41.46 [c]) | 22.50 (33.15 [c]) |
| $C_{44}$ | 32.50 (45.38 [c]) | 23.60 (40.46 [c]) | 9.40 (32.50 [c]) |
| *B* | 107.4 (137 ± 7 [b]; 124 [c]) | 100.0 (118 ± 4 [a]; 109 [c]) | 83.2 (84 ± 8 [a]; 91 [c]) |
| *G* | 56.00 | 45.99 | 28.35 |
| *G'* | 117.0 | 108.7 | 51.0 |
| *Y* | 143.17 | 119.63 | 76.37 |
| ν | 0.2782 | 0.3007 | 0.3470 |
| *G/B* | 0.521 | 0.460 | 0.341 |
| *CP* | - 2.9 | 4.0 | 13.1 |
| *A* | 0.28 | 0.22 | 0.10 |
| Phase | Th$_3$P$_4$ | Th$_3$As$_4$ | Th$_3$Sb$_4$ |
| *a* | 8.6471 (8.600-8.653 [a]) | 8.9001 (8.842-8.945 [a]) | 9.4514 (9.366-9.384 [a]) |
| $\rho^{theor}$ | 8.424 (8.56-8.44 [a]) | 9.382 (9.56 [a]) | 9.308 |
| $C_{11}$ | 193.30 | 175.00 | 141.30 |
| $C_{12}$ | 70.00 | 63.30 | 47.10 |
| $C_{44}$ | 55.50 | 48.60 | 37.70 |
| *B* | 111.0 | 100.5 | 78.5 |
| *G* | 57.88 | 51.38 | 41.22 |
| *G'* | 61.7 | 55.9 | 47.1 |
| *Y* | 147.96 | 131.71 | 105.23 |
| ν | 0.2780 | 0.2817 | 0.2766 |
| *G/B* | 0.521 | 0.516 | 0.525 |
| *CP* | 14.5 | 14.7 | 9.4 |
| *A* | 0.90 | 0.87 | 0.80 |

[a]Reference 5
[b]Reference 11
[c]Rerefence 15
* depending on Th/P stoichiometry (ThP$_x$; 0.55 ≤ x ≤ 0.96)

**Table 2**.

Total $N(E_F)$ and partial densities of states at the Fermi level (in states/eV atom), electronic heat capacity ($\gamma$, in mJ K$^{-2}$ mol$^{-1}$), and molar Pauli paramagnetic susceptibility ($\chi$, in $10^{-4}$ emu/mol) for metallic-like thorium *mono*-pnictides Th*Pn*. Available experimental data are given in parentheses.

| Phase | ThP | ThAs | ThSb |
|---|---|---|---|
| $N(E_F)$ | 1.38 (1.18-1.84 [a]) | 1.59 | 1.85 |
| Th 6*d* | 0.19 | 0.19 | 0.21 |
| Th 5*f* | 0.29 | 0.36 | 0.47 |
| *Pn* p | 0.05 | 0.07 | 0.03 |
| $\gamma$ | 3.25 (2.89-2.90 [a]) | 3.75 | 4.36 |
| $\chi$ | 0.44 | 0.51 | 0.60 |

[a]Reference 5

**Table 3**.

Calculated values of bandwidths (in eV) for cubic thorium pnictides Th*Pn* and Th$_3$*Pn*$_4$ (*Pn* = P, As, and Sb).

| Band types*/ phase | ThP | ThAs | ThSb | Th$_3$P$_4$ | Th$_3$As$_4$ | Th$_3$Sb$_4$ |
|---|---|---|---|---|---|---|
| 1 | 11.2 | 11.6 | 10.4 | 10.9 | 11.4 | 10.5 |
| 2 | 2.3 | 2.0 | 1.6 | 2.6 | 2.3 | 2.5 |
| 3 | 3.6 | 4.4 | 3.7 | 3.7 | 4.5 | 3.4 |
| 4 | 5.3 | 5.2 | 5.1 | 4.6 | 4.6 | 4.6 |

* 1- Full valence band (*Pn* s – E$_F$); 2- *Pn* s band; 3 - Gap (between *Pn* s - hybridized *Pn* p + Th (*p,d,f*) bands); 4 - hybridized *Pn* p + Th (*p,d,f*) band – up to E$_F$.

**Table 4.**
Calculated band gaps (direct (Γ-Γ) and indirect (Δ - Γ) transitions) for cubic thorium pnictides $Th_3P_4$ and $Th_3As_4$ in comparison with available experimental data.

| Gap/Potential | $Th_3P_4$ | $Th_3As_4$ |
|---|---|---|
| Direct Γ-Γ (DFT-GGA) | 0.44 | 0.51 |
| Indirect Δ – Γ (DFT-GGA) | 0.17 (0.003 [a]) | 0.28 (0.05 [a]) |
| Direct Γ-Γ (DFT-GGA+HF) | 0.64 | 0.66 |
| Indirect Δ – Γ (DFT-GGA+HF) | 0.40 | 0.44 |
| Experimental | 0.44 [b,c] | 0.39, [c] 0.43 [b] |

[a] Available calculated data (LDA approximation) are given in parentheses, Reference 12
[b] Reference 5
[c] Reference 43

**Table 5.**
Calculated Bader charge and effective atomic charges ($Q^B$ and $\Delta Q^{eff}$, in $e$) for cubic thorium pnictides Th$Pn$ and $Th_3Pn_4$ ($Pn$ = P, As, and Sb).

| phase | ThP | ThAs | ThSb | $Th_3P_4$ | $Th_3As_4$ | $Th_3Sb_4$ |
|---|---|---|---|---|---|---|
| Th ($Q^B$) | 10.30 | 10.40 | 10.53 | 110.12 | 10.22 | 10.44 |
| Th ($\Delta Q^{eff}$) | + 1.70 | +1.60 | +1.47 | +1.88 | +1.78 | +1.56 |
| Pn ($Q^B$) | 6.70 | 6.60 | 6.47 | 6.41 | 6.33 | 6.19 |
| Pn ($\Delta Q^{eff}$) | -1.70 | -1.60 | -1.47 | -1.41 | -1.33 | -1.19 |

**FIGURES**

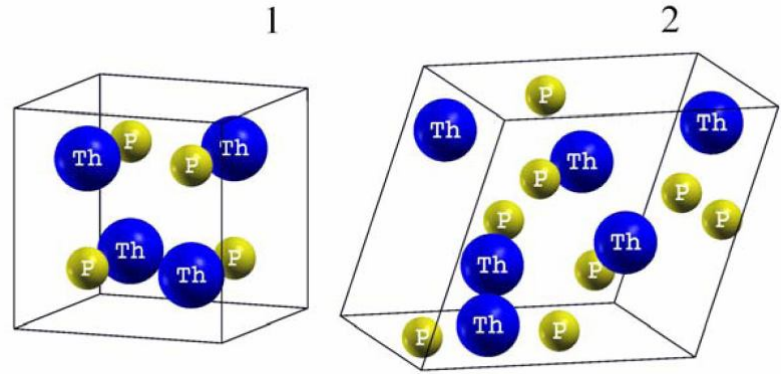

**Figure 1**. (*Color online*) Crystal structures of cubic ThP (1) and Th$_3$P$_4$ (2).

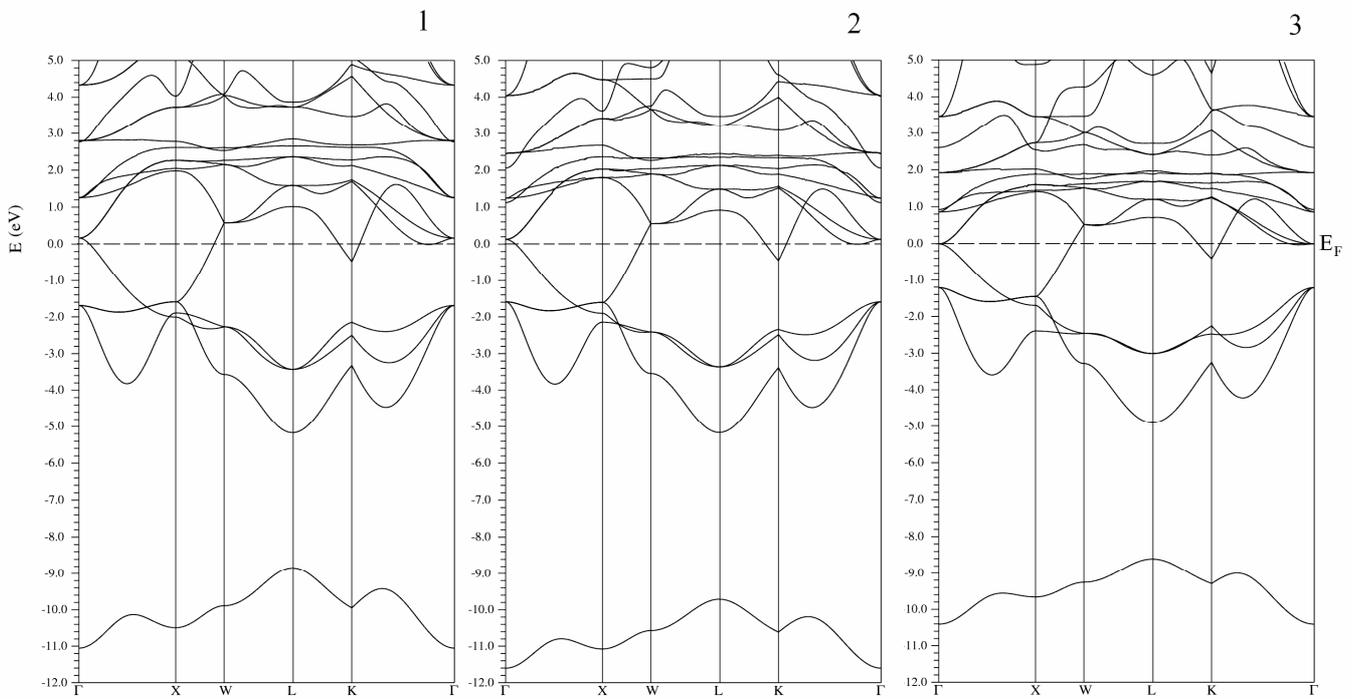

**Figure 2**. Band structures of ThP (1), ThAs (2), and ThSb (3).

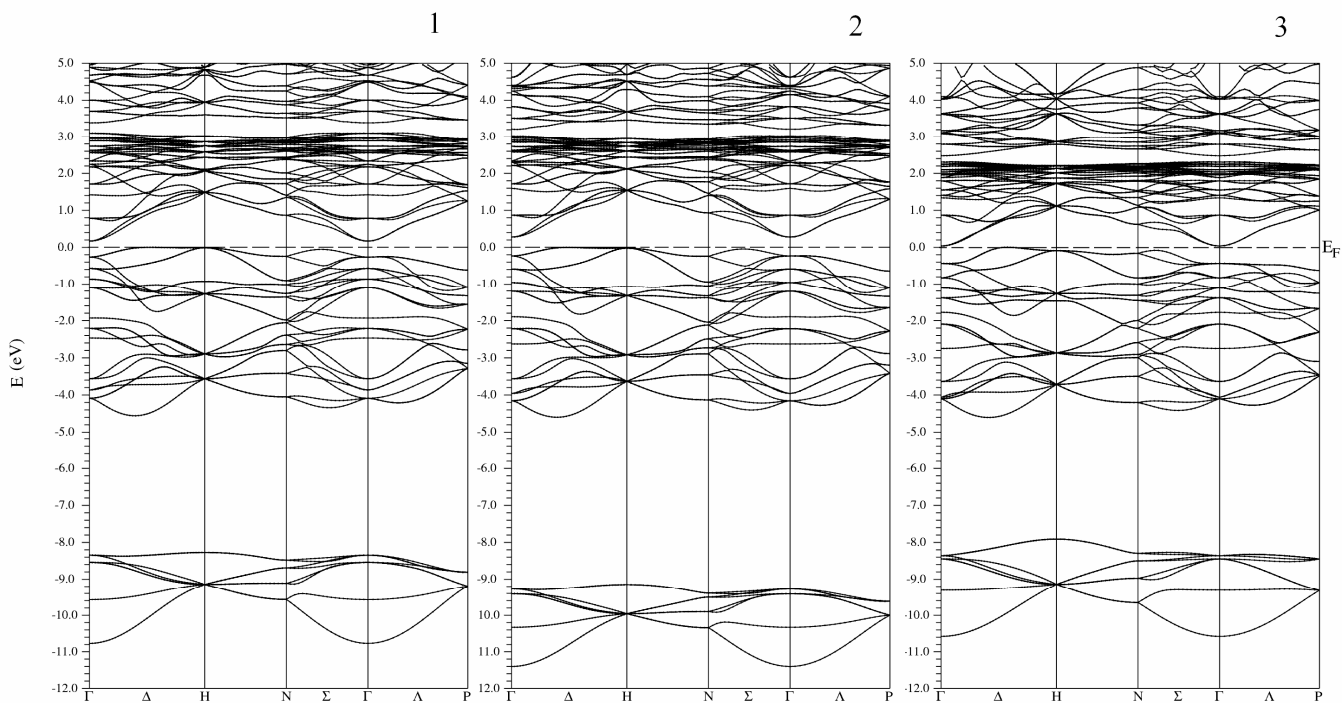

**Figure 3.** Band structures of $Th_3P_4$ (1), $Th_3As_4$ (2), and $Th_3Sb_4$ (3).

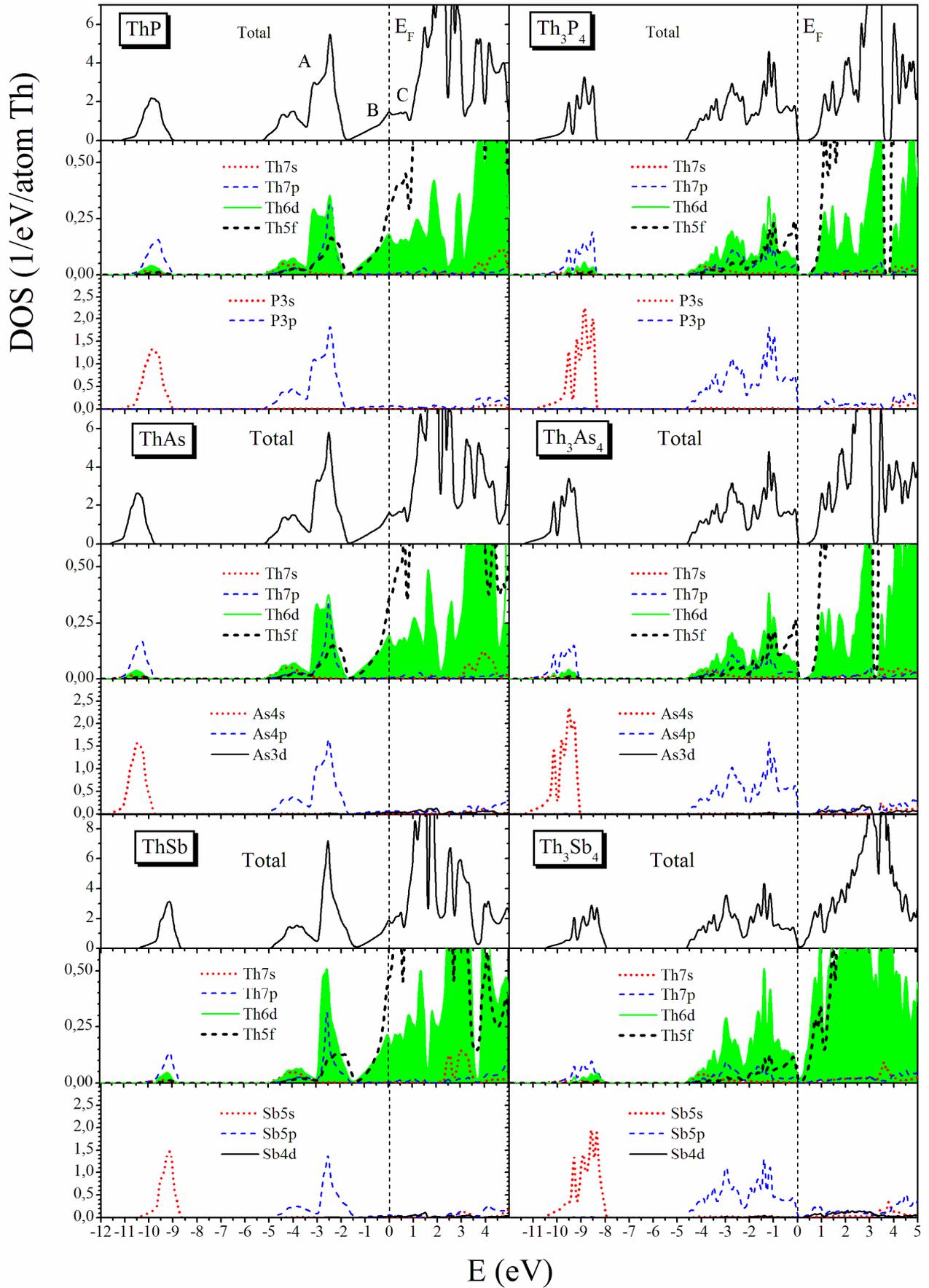

**Figure 4.** (*Color online*) Total and partial density of states for cubic thorium pnictides Th*Pn* and Th$_3$*Pn*$_4$ (*Pn* = P, As, and Sb).

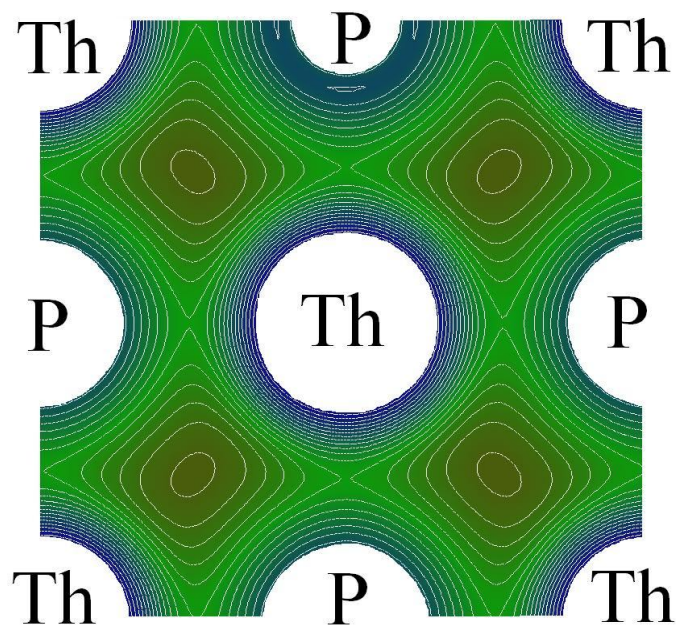

**Figure 5.** (*Color online*) Charge density map in the (100) plane of B1-like ThP. The interval between the iso-lines is 0.1 electrons/Å$^3$.

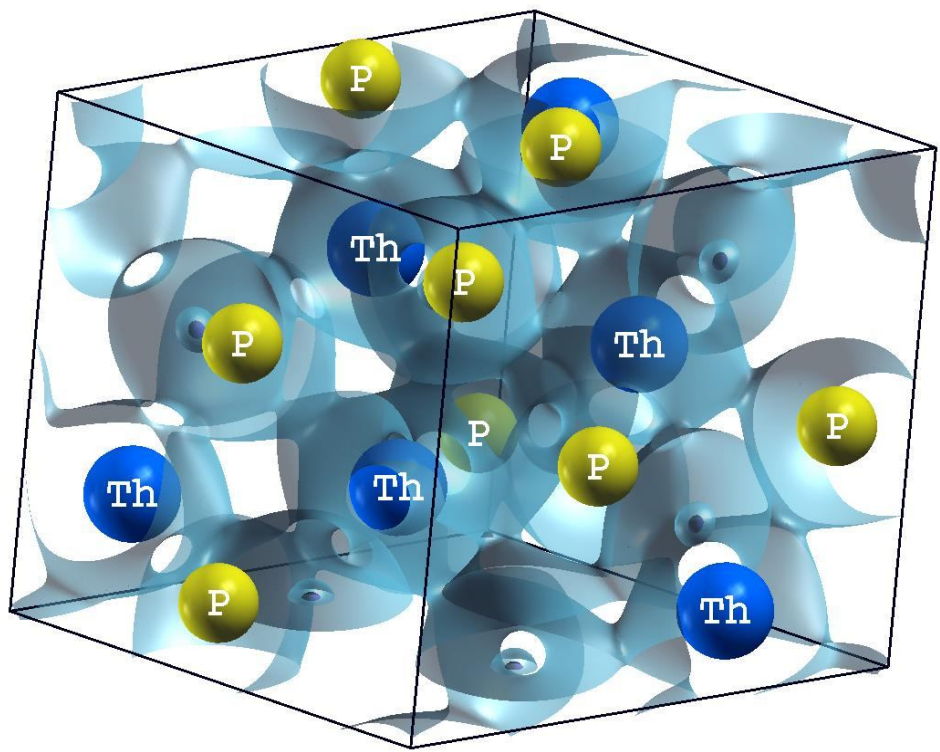

**Figure 6.** (*Color online*) Isoelectronic surface ($\rho = 3.5$ electrons/Å$^3$) of the charge density in cubic Th$_3$P$_4$, which illustrates the intra-atomic bonding for *tetra*-pnictides Th$_3$Pn$_4$.